\documentclass[onecolumn,showpacs,preprint]{revtex4}

\usepackage{bbm}
\usepackage{mathrsfs}
\usepackage{graphicx}
\usepackage{dcolumn}
\usepackage{array}
\usepackage{bm}
\usepackage{amsmath}
\usepackage{amsfonts}
\usepackage{amssymb}

\input{tcilatex}

\begin{document}

\title{Anisotropic spin transport in two-terminal mesoscopic rings: the
Rashba and Dresselhaus spin-orbit interactions}
\author{Miao Wang}
\author{Kai Chang}
\email{kchang@red.semi.ac.cn}
\affiliation{SKLSM, Institute of Semiconductors, Chinese Academy of Sciences, P. O. Box
912, Beijing 100083, China}

\begin{abstract}
We investigate theoretically the spin transport in two-terminal mesoscopic
rings in the presence of both the Rashba spin-orbit interaction (RSOI) and
the Dresselhaus spin-orbit interaction (DSOI). We find that the interplay
between the RSOI and DSOI breaks the original cylindric symmetry of
mesoscopic ring and consequently leads to the anisotropic spin transport,
i.e., the conductance is sensitive to the positions of the incoming and
outgoing leads. The anisotropic spin transport can survive even in the
presence of disorder caused by impurity elastic scattering in a realistic
system.
\end{abstract}

\maketitle

\section{Introduction}

In recent years, the spin-orbit interaction (SOI) in low-dimensional
semiconductor structures has attracted considerable attention because of its
potential application in all-electrical controlled spintronic devices.~\cite%
{wolf,Tsitsishvili} There are two types of SOI in conventional
semiconductors. One is the Rashba spin-orbit interaction(RSOI) induced by
structure inversion asymmetry,~\cite{Rashba,Bychkov} and the other is the
Dresselhaus spin-orbit interaction(DSOI) induced by bulk inversion asymmetry~%
\cite{Dresselhaus}. The strength of the RSOI can be tuned by external gate
voltages or asymmetric doping. In thin quantum wells, the strength of the
DSOI is comparable to that of the RSOI.~\cite{Lommer} The interplay between
the RSOI and DSOI leads to a significant change in the transport property.
There are a few works on the effects of the competition between these two
types of SOI on the transport properties of 2DEG,~\cite{Ganichev,Chang,Yang}%
especially in mesoscopic rings~\cite{Vasilo}. The circular photogalvanic
effect can be used to separate the contribution of the RSOI and DSOI, and
the relative strengths of the RSOI and DSOI can be extracted from the
photocurrent.~\cite{Ganichev} The RSOI and DSOI can interfere in such a way
that the spin dependent features disappear even though the individual SOI is
still strong, e.g., vanishing spin splitting in the presence of the
equal-strength RSOI and DSOI.~\cite{Ganichev} This cancellation results in
extremely long spin relaxation time in specific crystallographic directions,
and the disappearance of the beating pattern in SdH oscillation.~\cite{Yang}

Recently, advanced growth techniques have made it possible to fabricate high
quality semiconductor rings,~\cite{Fuhrer} which have attracted considerable
attention due to the intriguing quantum interference phenomenon arising from
their unique topological geometry. The Aharonov-Bohm (AB) and the
Aharonov-Casher (AC) effects are typical examples of quantum mechanical
phase interference, which have been demonstrated experimentally~\cite%
{Tonomura,Kong} and theoretically~\cite{Balatsky} on semiconductor rings.
The quantum transport properties through semiconductor ring structures with
the RSOI alone have attracted considerable interest.~\cite%
{Molnar,SoumaBK,Entin,Cao,Foldi,Nitta} SOIs in semiconductors behave like an
in-plane momentum-dependent magnetic field and lead to a lifting of spin
degeneracy of energy bands. This effective magnetic field induces a wave
phase difference between the upper arm and lower arm, resulting in the
oscillation of the conductance.~\cite{wolf,Nitta,Diego} Therefore, the
conductance oscillates with increasing the strength of the RSOI.\cite%
{Molnar,SoumaBK} The ring subjected to the DSOI alone shows the exact same
oscillation, since the Hamiltonian of the RSOI alone is mathematically
equivalent to that of the DSOI alone by a unitary transformation.~\cite%
{Sheng} The interplay between the RSOI and DSOI results in a periodic
potential in an isolated ring, producing the gap in the energy spectrum,
suppressing the persistent currents,~\cite{Sheng} and breaking the
cylindrical symmetry of mesoscopic rings. This interesting feature leads to
the anisotropic spin transport and could be detected using the transport
property in an open two-terminal mesoscopic ring. This anisotropic spin
transport is a new result, is dominant difference between our work and the
previous studies,~\cite{Molnar,SoumaBK,Entin,Cao,Foldi} and should be
important for the potential application of spintronic devices.

In this paper, we investigate theoretically the spin transport in
two-terminal mesoscopic rings in the presence of both the RSOI and DSOI. We
find that the interplay between the RSOI and DSOI leads to a significant
change in the transmission, the localization of electrons, and the spin
polarization of the current. This interplay weakens and smoothens the
oscillation of the conductance, and breaks the original cylindrical
symmetry, leading to the anisotropic spin transport. The paper is organized
as follows, in Sec.~\ref{sec:theory}, we present the theoretical model and
formulation. The numerical results and discussions are given in Sec.~\ref%
{sec:results}. Finally, the conclusion is given in Sec.~\ref{sec:conclusions}

\section{\label{sec:theory}THEORETICAL MODEL}

\begin{figure}[ptb]
\includegraphics[width=\columnwidth]{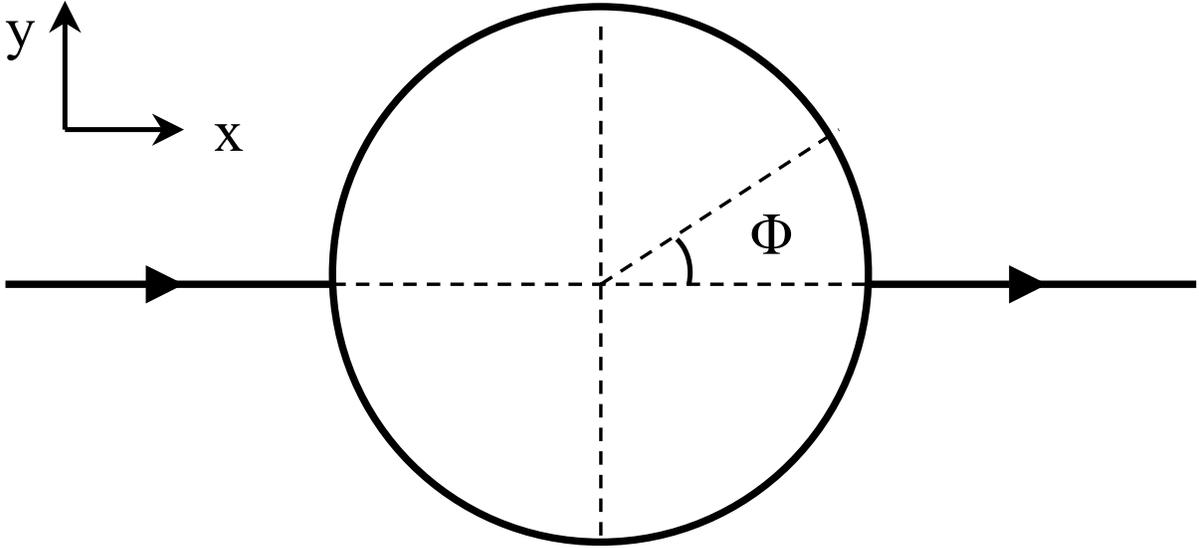}
\caption{Schematic diagram of a 1D semiconductor mesoscopic ring with two
leads. Electrons are injected from the left lead, pass through the ring, and
exit from the right lead. SOI only exists in the ring.}
\label{fig:schematic}
\end{figure}
A semiconductor mesoscopic ring (see Fig.~\ref{fig:schematic}) in the
presence of the RSOI and DSOI can be described by the single-particle
effective mass Hamiltonian
\begin{align}
\hat{H} & = \frac{-\hbar^{2}k^{2}}{2m^{\ast}}+\alpha(\sigma_{x}k_{y}-%
\sigma_{y}k_{x}){}  \notag \\
& +\beta(\sigma_{x}k_{x}-\sigma_{y}k_{y})+V(r),  \label{hami}
\end{align}
where the $x$ axis is along the [100] direction, $k = -i\nabla$ is the
electron wave vector, $m^{\ast}$ is the electron effective mass, $%
\sigma_{i}(i=x, y, z)$ are the Pauli matrices, $\alpha$ is the strength of
the RSOI, and $\beta$ is the strength of the DSOI. $V(r)$ is the radial
confining potential, which is neglected hereafter since we consider that
electrons only occupy the lowest subband in a ring with narrow width. The
one-dimensional Hamiltonian of a ring in a dimensionless form in lattice
representation is~\cite{Souma}
\begin{align}
\hat{H}_{ring} &
=\sum^{N}_{n=1}\sum_{\sigma=\uparrow,\downarrow}\varepsilon_{n}\hat{c}%
^{\dag}_{n,\sigma}\hat{c}_{n,\sigma}  \notag \\
&-\sum^{N}_{n=1}\sum_{\sigma,\sigma^{^{\prime
}}=\uparrow,\downarrow}[t^{n,n+1;\sigma,\sigma^{^{\prime }}}_{\phi}\hat{c}%
^{\dag}_{n;\sigma}\hat{c}_{n+1;\sigma^{^{\prime }}} + h.c.],
\end{align}
where the hopping energies are given in the $2\times2$ matrix form as:
\begin{align}
t^{n,n+1}_{\phi} & =t\hat{I}_{s}-i\frac{\alpha}{2a}(cos\phi_{n,n+1}%
\sigma_{x}+sin\phi_{n,n+1}\sigma_{y})  \notag \\
&-i\frac{\beta}{2a}(cos\phi_{n,n+1}\sigma_{y}-sin\phi_{n,n+1}\sigma_{x}),
\end{align}
where $\phi$ is the angular coordinate and $\varepsilon_{n}$ is the on-site
potential energy. The operator $\hat{c}_{n,\sigma} (\hat{c}%
^{\dag}_{n,\sigma})$ annihilates (creates) a spin $\sigma$ electron at the
site $n$ of the ring. $\phi_{n,n+1}$ is the angle between the $n$-th site
and the $n+1$-th site. $t=\hbar^{2}/2m^{\ast}a^{2}$, with $a$ being the
lattice spacing constant, is the nearest-neighbor hopping term in the lead.

The spin-resolved conductance of a two-terminal device can be obtained by
using the Landauer-B\"{u}ttiker's formula~\cite{BK}:
\begin{equation}
\mathbf{G}=\left(
\begin{array}{cc}
G_{\uparrow \uparrow} & G_{\uparrow \downarrow} \\
G_{\downarrow \uparrow} & G_{\downarrow \downarrow}%
\end{array}
\right) =\frac{e^{2}}{h}\sum_{p,p^{^{\prime}}=1}^{M}\left(
\begin{array}{cc}
|\mathbf{t}_{pp^{^{\prime}},\uparrow \uparrow}|^{2} & |\mathbf{t}%
_{pp^{^{\prime}},\uparrow \downarrow}|^{2} \\
|\mathbf{t}_{pp^{^{\prime}},\downarrow \uparrow}|^{2} & |\mathbf{t}%
_{pp^{^{\prime}},\downarrow \downarrow}|^{2}%
\end{array}
\right),  \label{LB-G}
\end{equation}
where $M$ is the number of conducting channels, the transmission matrix
elements $\mathbf{t}=2\sqrt{-\text{Im}\sum^{r}_{L} \otimes I_{s}}\cdot
G^{r}_{1N}\cdot\sqrt{-\text{Im}\sum^{r}_{R} \otimes I_{s}}$ and $|\mathbf{t}%
_{nn^{^{\prime}}, \sigma \sigma^{^{\prime}}}|^{2}$ represents the
probability for a spin-$\sigma$ electron incoming from the left lead in the
orbital state $|n\rangle$ to appear as a spin-$\sigma^{^{\prime}}$ electron
in the orbital channel $|n^{^{\prime}}\rangle$ in the right lead.

We can calculate the conductance from lead $p$ to lead $q$ by using the
Fisher-Lee relation~\cite{Fisher}. The detailed formula can be found in the
Ref.~\onlinecite{Green}:
\begin{equation}  \label{FL-G}
\mathbf{G}^{R}= [EI-H_{c}-\Sigma^{R}]^{-1},
\end{equation}
\begin{equation}  \label{FL-Tr}
\overline{\mathbf{T}}_{pq}= Tr[\Gamma_{p}G^{R}\Gamma_{q}G^{A}],
\end{equation}
where $H_{c}$ is the Hamiltonian of the 1D isolated ring. $%
\Gamma_{p}(i,j)=\sum_{m}\chi_{m}(p_{i}) \frac{\hbar v_{m}}{a}\chi_{m}(p_{j})$
describes the coupling of the ring conductor to the leads. We assume the
RSOI and DSOI only exist in the ring, and are absent in the leads. The
self-energy $\Sigma^{R}= \sum_{p=1,2}\Sigma_{p}^{R}$, where $%
\Sigma_{p}^{R}(i,j)= t^{2}g_{p}^{R}(p_{i},p_{j})$, describes the effect of
the external leads on the ring. The Green's function between two points
along the leads is given by $g_{p}^{R}(p_{i},p_{j})=-\frac {1}{t}%
\sum_{m}\chi_{m}(p_{i})exp[ik_{m}a] \chi_{m}(p_{j})$. The function $%
\chi_{m}(p_{i})$ describes the $m$-th mode in lead $i$. In this paper, we
take $a$ as the length unit and $E_{0}=\hbar^{2}/2m^{\ast}a^{2}$ as the
energy unit.

The local density of electron states is~\cite{Green}:
\begin{equation}  \label{density}
\rho(r,E)=\frac{1}{2\pi}A(r,r;E)=-\frac{1}{\pi }\mathbf{Im} [G^{R}(r,r;E)],
\end{equation}
where $A\equiv i[G^{R}-G^{A}]$ is the spectral function, which can also be
written:
\begin{align}  \label{den}
\rho(r,E) &\sim\sum_{n}\frac{1}{2\pi}\frac{\gamma_{n}\psi_{n}(r)%
\phi^{*}_{n}(r)}{(E-\varepsilon_{n0}+\Delta_{n})^{2}+(\gamma_{n}/2)^{2}}
\notag \\
&\rightarrow\sum_{n}\delta(E-\varepsilon_{n0})|\psi_{n}(r)|^{2} \quad
\mathbf{as} \quad\gamma_{n}\rightarrow0,
\end{align}
where $\hbar/2\gamma_{n}$ represents the lifetime of an electron remaining
in state $n$ before it escapes into the leads, $\varepsilon_{n0}$ is the
eigenenergy of the isolated conductor, and $\psi$ ($\phi$) is the
eigenstates of the effective Hamiltonian [$H_{c}+\Sigma^{R}$] ([$%
H_{c}+\Sigma^{A}$])~\cite{Green}.

\section{\label{sec:results}RESULTS AND DISCUSSIONS}

\subsection{\label{both}1D ring with both RSOI and DSOI}

Many previous works investigating the spin transport through a 1D ring
account only for the RSOI.~\cite{Molnar} The RSOI behaves like an effective
in-plane momentum-dependent magnetic field. This effective magnetic field
induces a phase difference between the electrons traveling clockwise and
counterclockwise along the ring's upper and lower arms. Therefore, the
conductance of a 1D ring in the presence of the RSOI oscillates
quasi-periodically with changing the strength of the RSOI and the Fermi
energy $E_{F}$.

\begin{figure}[ptb]
\includegraphics[width=\columnwidth]{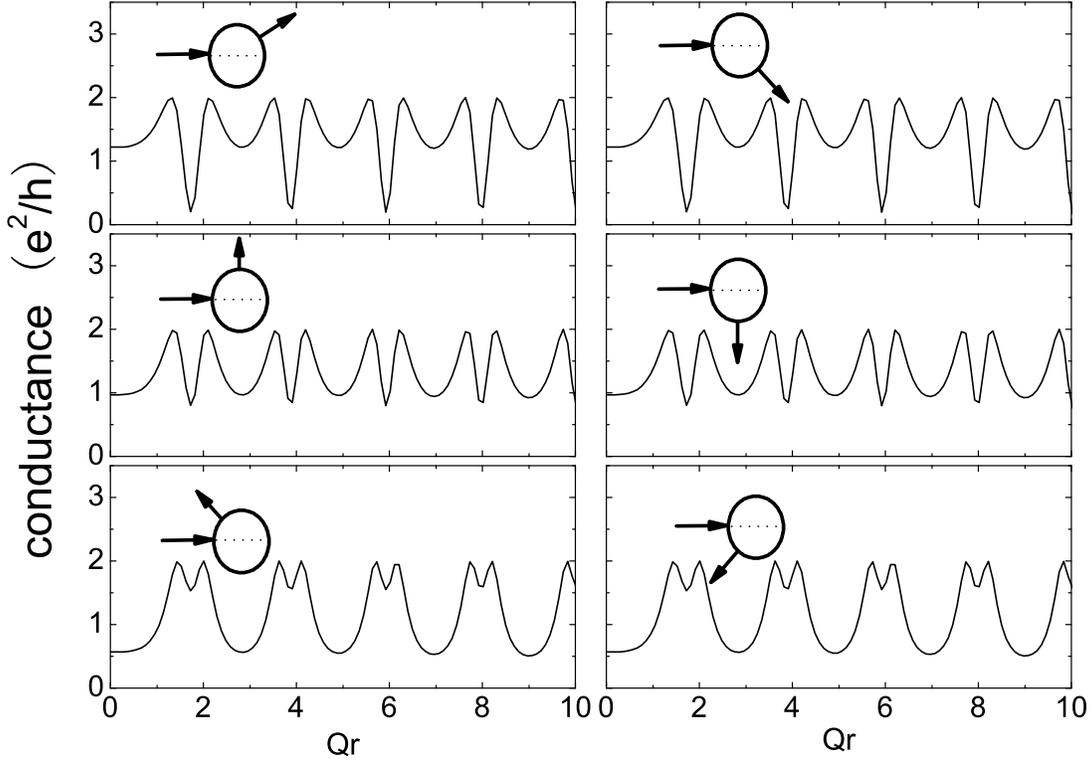}
\caption{ The conductance through a 1D ring in the presence of the RSOI or
DSOI alone as a function of the strength of the RSOI $Q_{r}\equiv \protect%
\alpha N/2ta \protect\pi$, $E_{F}$=-0.1, and the outgoing lead is located at
$\pm \frac{1}{2}\protect\pi,\pm \frac{1}{4}\protect\pi,\pm \frac{3}{4}%
\protect\pi$, respectively (see the insets).}
\label{fig:RSOI}
\end{figure}

We study the transport through a mesoscopic ring in the presence of both the
RSOI and DSOI. First, we consider the ballistic transport through the
mesoscopic ring in the presence of the RSOI(DSOI) alone. In Fig.~\ref%
{fig:RSOI}, we plot the conductance through a 1D ring as a function of the
strength of the RSOI $Q_{r}$. This figure shows that the conductances are
exactly same when the right lead is located at symmetric positions, e.g., $%
\phi=\pm\frac{1}{4}\pi, \pm\frac{1}{2}\pi$, and $\pm\frac{3}{4}\pi$. The
RSOI or DSOI alone in the ring does not break the cylindrical symmetry and
the transport is still isotropic when the outgoing leads are located at
symmetric positions with respect to the $x$-axis (see the dashed lines in
the insets of Fig.~\ref{fig:RSOI}). The quantum interference between the
alternation paths, the spin-up or spin-down clockwise and anticlockwise, is
responsible for the oscillation of the conductance.

\begin{figure}[ptb]
\includegraphics[width=\columnwidth]{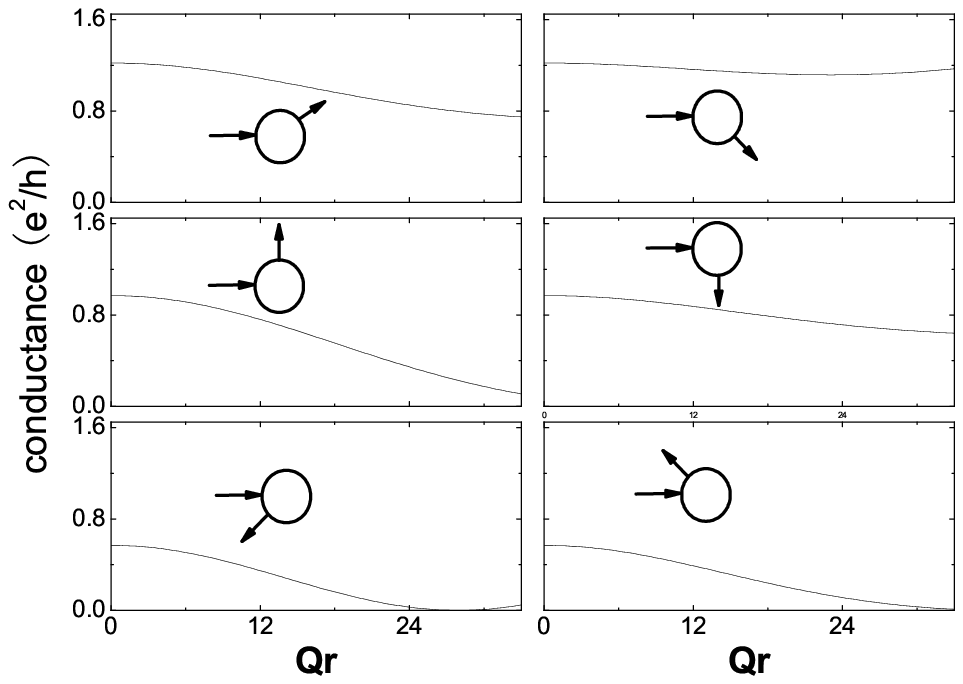}
\caption{ The conductance through a 1D ring in the presence of the RSOI and
DSOI as a function of the strength of the RSOI and DSOI, $Q_{d}\equiv
\protect\beta N/2ta \protect\pi$, $Q_{r}=Q_{d}$, $E_{F}$=-0.1. The outgoing
lead is located at $\pm\frac{1}{2}\protect\pi, \pm\frac{1}{4}\protect\pi, \pm%
\frac{3}{4}\protect\pi$ respectively.}
\label{fig:RandD}
\end{figure}

When the 1D mesoscopic ring is subjected to both the RSOI and DSOI, as shown
in Fig.~\ref{fig:RandD}, the conductances become asymmetric when the
outgoing lead is located at symmetric positions, e.g., $\phi=\pm\frac{1}{4}%
\pi, \pm\frac{1}{2}\pi$, and $\pm\frac{3}{4}\pi$. The anisotropy of the
conductance is induced by the interplay of the RSOI and DSOI, which leads to
a periodic potential $\frac{\alpha\beta}{2}\sin{2\phi}$.~\cite{Sheng} The
height of the periodic potential is determined by the product of the
strengths of the RSOI and DSOI, and the periodicity of the potential is
fixed at $\pi$. The potential exhibits barriers at $\phi=\frac{1}{4}\pi, -%
\frac{3}{4}\pi$, and the valleys at $\phi=-\frac{1}{4}\pi, \frac{3}{4}\pi$.
Thus, the conductance displays asymmetric features for the symmetric
positions of the outgoing leads.

\begin{figure}[ptb]
\includegraphics[width=\columnwidth]{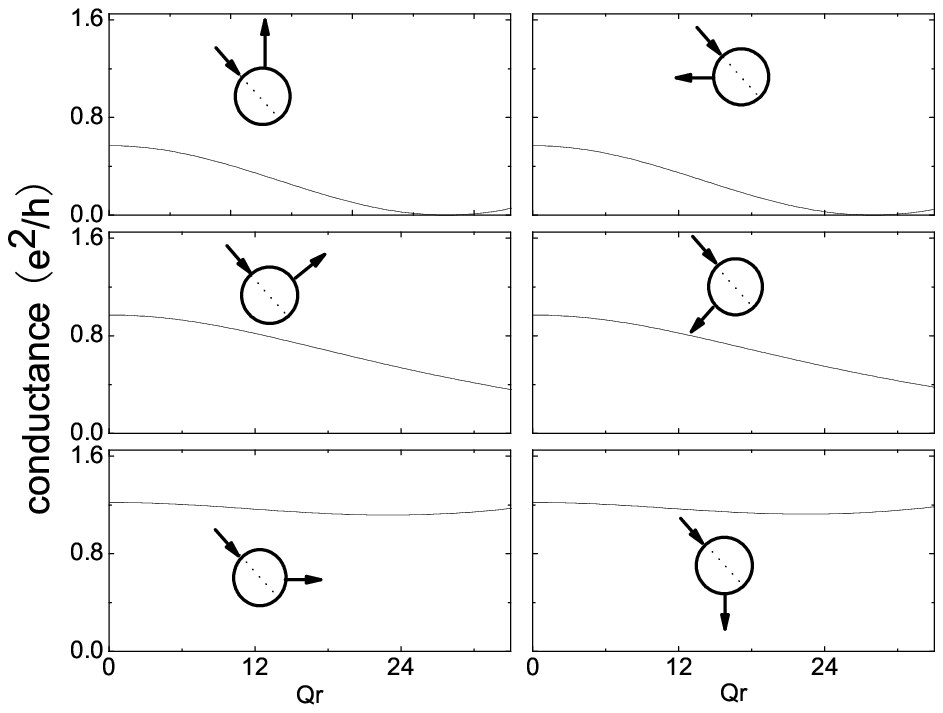}
\caption{ Same as Fig.~\protect\ref{fig:RandD}, but the incoming lead is
located at $\protect\phi=\frac{3}{4}\protect\pi$, and the outgoing lead is
located at $\protect\phi=0, \pm\frac{1}{2}\protect\pi, \frac{1}{4}\protect\pi%
, -\frac{3}{4}\protect\pi, \protect\pi$, respectively.}
\label{fig:newaxis}
\end{figure}

If the incoming lead locates at $\phi=\frac{3}{4}\pi$ (see Fig.~\ref%
{fig:newaxis}), we find the transmission becomes symmetric for the outgoing
lead locating at the symmetric positions respect to the new incoming lead.
In Fig.~\ref{fig:newaxis}, we plot the conductance of a 1D ring with the
incoming lead located at $\phi=\frac{3}{4}\pi$. The conductance becomes
symmetric again with respect to the straight line $\phi=\frac{3}{4}\pi$ and $%
\phi=-\frac{1}{4}\pi$ (the dashed lines in the insets of Fig.~\ref%
{fig:newaxis}). The periodic potential $\frac{\alpha\beta}{2}\sin{2\phi}$
induced by the interplay between the RSOI and DSOI~\cite{Sheng} results in
the maxima at $\phi=\frac{1}{4}\pi, -\frac{3}{4}\pi$, and the minima at $%
\phi=\frac{3}{4}\pi, -\frac{1}{4}\pi$.

In order to describe the magnitude of the anisotropy of the conductance
induced by the interplay of the RSOI and DSOI, we define the ratio $\eta$
as:
\begin{equation}  \label{Tr}
\eta(\phi,-\phi)=\frac{G_{\phi}-G_{-\phi}}{(G_{\phi}+G_{-\phi})/2}\ ,
\end{equation}
where $G_{\pm\phi}$ is the conductance when the right lead is located at the
positions with an angle $\pm\phi$ with respect to the $x$ axis.

\begin{figure}[ptb]
\includegraphics[width=\columnwidth]{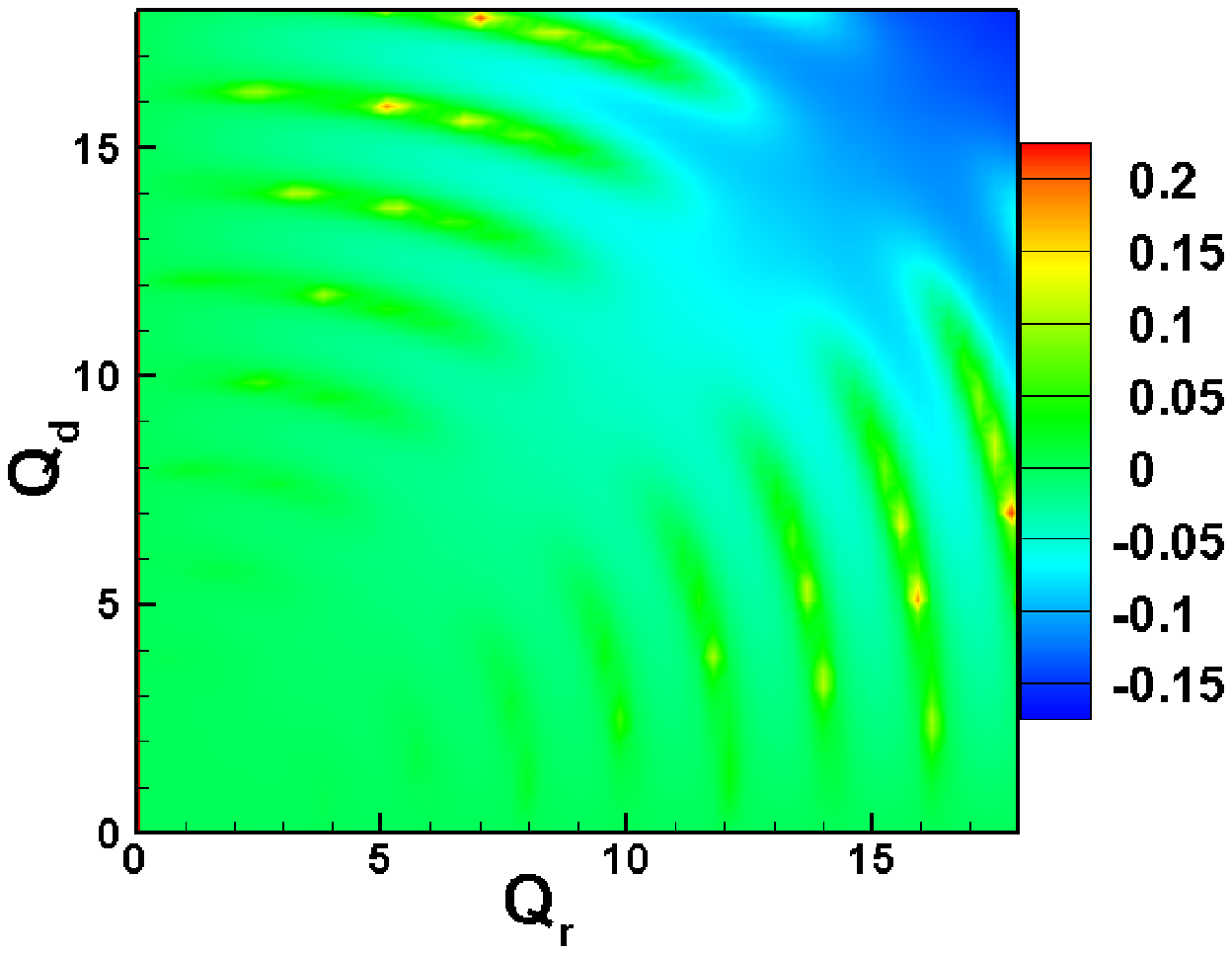}
\caption{ (Color online) The ratio $\protect\eta$ as a function of the
strength of the RSOI $Q_{r}$ and DSOI $Q_{d}$, when $E_{F}=-0.1$. The
incoming lead is located at $\protect\phi=\protect\pi$, and the outgoing
lead is located at $\protect\phi= \protect\pi/4, -\protect\pi/4$,
respectively.}
\label{fig:Tr}
\end{figure}

\begin{figure}[tbp]
\includegraphics[width=\columnwidth]{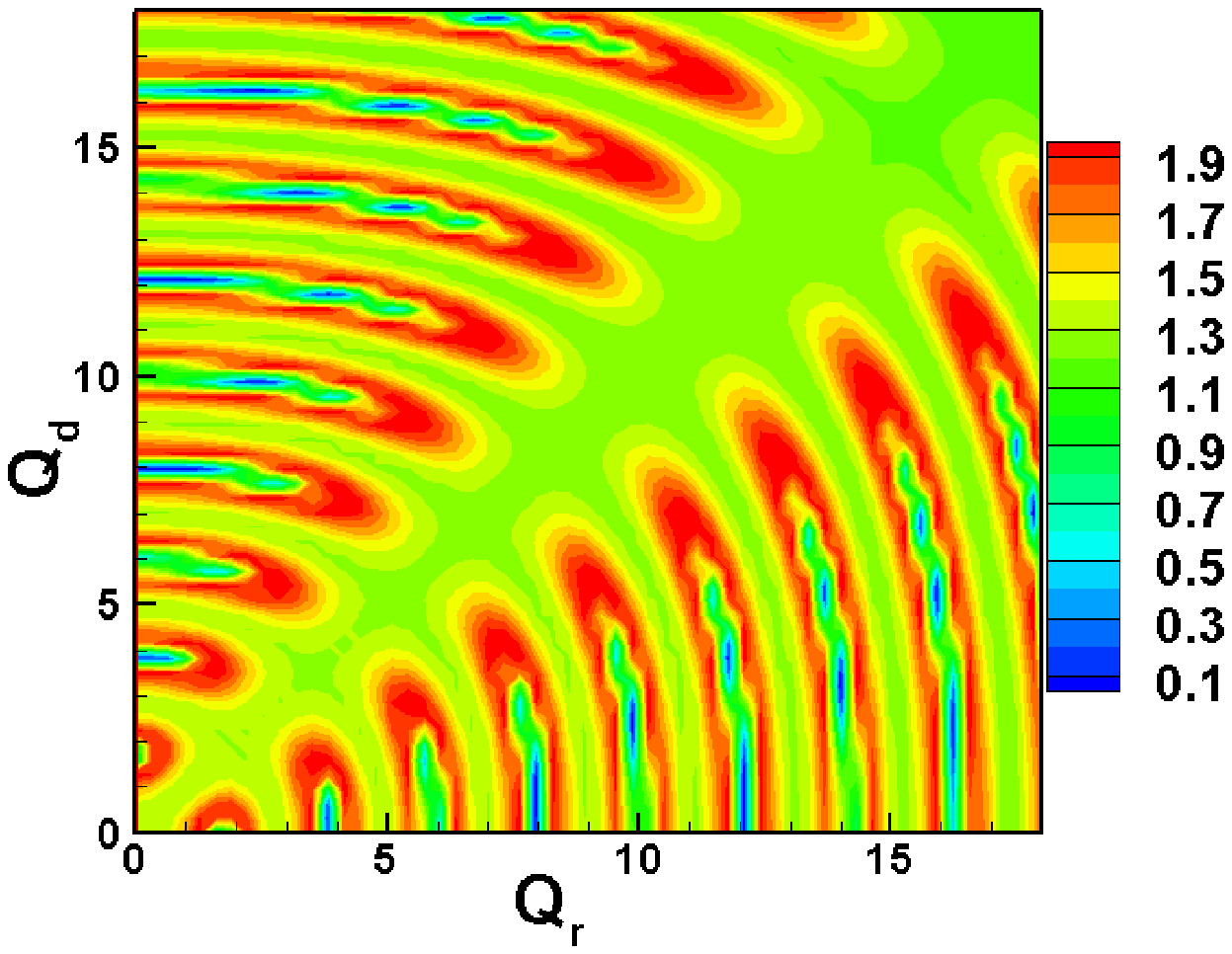}
\caption{ (Color online) The conductance of a 1D ring as a function of the
strength of the RSOI $Q_{r}$ and DSOI $Q_{d}$, when $E_{F}=-0.1$. The
incoming lead is located at $\protect\phi =\protect\pi $, while the outgoing
lead is located at $\protect\phi =0$.}
\label{fig:ab}
\end{figure}

In Fig.~\ref{fig:Tr}, we plot $\eta(\pi/4, -\pi/4)$ as a function of the
strength of the RSOI and DSOI when the left lead is located at the position
of $\phi= \pi$. $\eta$ oscillates with the changing strength of the RSOI and
DSOI. The maximum of the anisotropy of the conductance can approach $20\%$.
This anisotropic transport can be interpreted as follows. The interplay
between the RSOI and DSOI leads to an effective periodic potential $\frac{%
\alpha\beta}{2}\sin{2\phi}$.~\cite{Sheng} The potential height is related to
the strength of the RSOI and DSOI. $\eta(\phi, -\phi)= 0$ when the ring
subjected to the DSOI alone because the periodic potential $\frac{\alpha\beta%
}{2}\sin{2\phi}$ disappears when $\alpha=0$. This effective periodic
potential exhibits the maxima at $\phi=\frac{1}{4}\pi$, and $-\frac{3}{4}\pi$%
, and the minima at $\phi=-\frac{1}{4}\pi$, and $\frac{3}{4}\pi$. Therefore,
the interplay between RSOI and DSOI breaks the cylindrical symmetry of the
ring (see Fig.~\ref{fig:ab}).

\begin{figure}[tbp]
\includegraphics[width=\columnwidth]{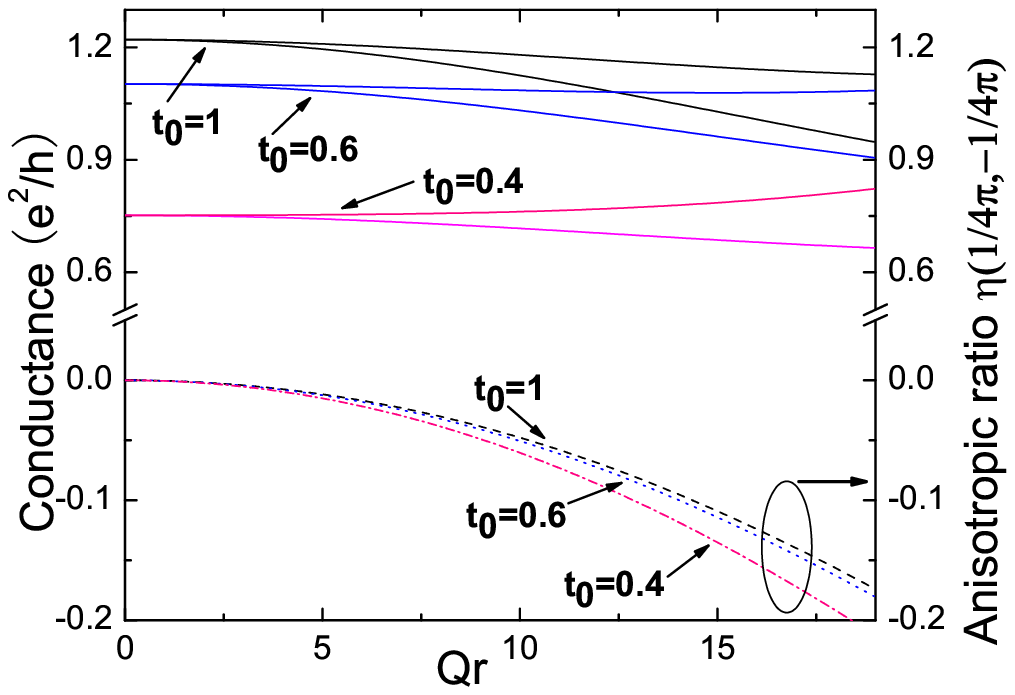}
\caption{ (Color online) The conductance of 1D ring as a function of the
strength of equal RSOI and DSOI, when $E_{F}=-0.1$ for different coupling
strengths $t_0=1, 0.6, 0.4$.}
\label{fig:change-t}
\end{figure}

In order to clarify the effect of the invasive role of the lead on the
anisotropy of the spin transport, we consider different strengths between
the ring and leads (as shown in Fig.~\ref{fig:change-t}). We find that the
conductance decreases with decreasing the coupling strength, but the
anisotropy ratios are almost same as before. We believe that the anisotropic
spin transport property is caused by the interplay between the Rashba and
Dresselhaus spin-orbit interactions.

Fig.~\ref{fig:ab} describes how the conductance varies with the variation of
the strengths of the RSOI and DSOI. The conductance oscillates
quasiperiodically as the strengths of the RSOI and DSOI increase, and is
symmetric with respect to the straight line $\alpha=\beta$, since the
Hamiltonian of the RSOI and that of the DSOI are equivalent and can be
transferred by the $SU(2)$ unitary transformation. The contribution from the
RSOI and DSOI to the spin splitting of electrons cancel each other,~\cite%
{Sheng} which results in the disappearance of the oscillation along $%
\alpha=\beta$. This feature provides a possible way to detect the strength
of the DSOI since the strength of the RSOI can be tuned by the external
electric fields.

\begin{figure}[ptb]
\includegraphics[width=\columnwidth]{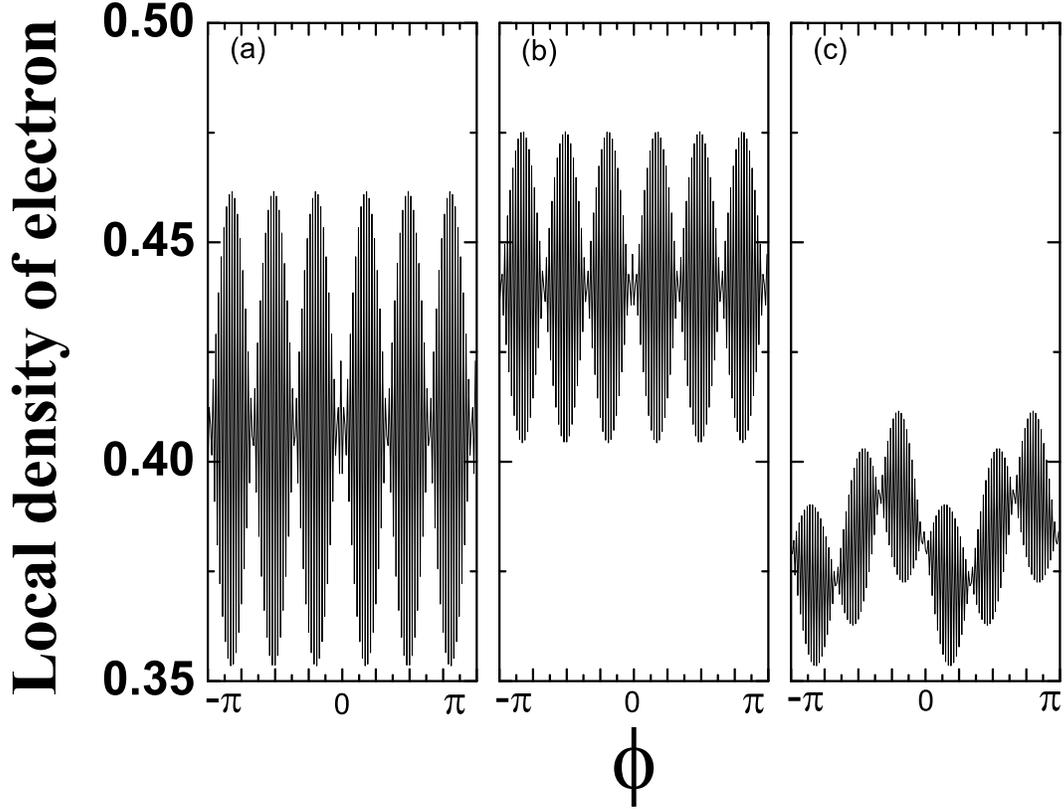}
\caption{The local density of electrons along the ring $\protect\phi$ when $%
E_{F}=0.1$ (a) without the RSOI and DSOI; (b) with the RSOI alone; (c) with
equal RSOI and DSOI ($Q_{r}=Q_{d}=11.3$).}
\label{fig:density}
\end{figure}

Below, we demonstrate that the interplay between the RSOI and DSOI also
results in the variation of the local density of electrons in the ring. In
Fig.~\ref{fig:density}, we plot the local density of electrons in the ring
from Eq.~\ref{density} with and without the SOI. Fig.~\ref{fig:density}(a)
and (b), shows that the local density of electrons shows slow and very rapid
oscillations. The fast oscillation comes from the contribution of each site
of the lattice, while the slow variation of the envelope corresponds to the
bound (quasibound) states in the isolated (open) ring. This feature is
analogous to the situation of the effective mass theory, where the electron
wave function can be expressed as the product of two parts: the band-edge
Bloch function and the slow varying envelope function. The former denotes
the contribution from the atomic wave function, and the latter describes the
bound (quasibound) state from the external potential, e.g., the quantum well
potential. Similar results can be found in Ref.~\onlinecite{Li}.

There is only a slight difference between the local densities of electron
states with and without the RSOI, but a significant change in the presence
of both the RSOI and DSOI (see Fig.~\ref{fig:density}(c)). The local density
of electrons exhibit maxima at $\phi=-\frac{ 1}{4}\pi, \frac{3}{4}\pi$. This
characteristic is also caused by the periodic potential induced by the
interplay between the RSOI and DSOI. The positions of $\phi=\frac{1}{4}\pi, -%
\frac{3}{4}\pi$ ($\phi=-\frac{1}{4}\pi, \frac{3}{4}\pi$) correspond to a
potential barrier (well), where the local density of electron states is
smaller (larger). The interplay between the RSOI and DSOI induces periodic
potential and breaks the original cylindrical symmetry of the ring,
consequently changing the local density of electron states.

The above analysis assumes perfectly clean 1D systems, in which there is no
elastic or inelastic scattering at $T=0$. In a realistic system, there will
be many impurities in the sample. Disorder could be incorporated by the
fluctuation of the on-site energies, which distribute randomly within the
range width $w[\varepsilon_{n}\rightarrow \varepsilon_{n}+ w_{n}$ with $%
-w/2<w_{n}<w/2]$.

\begin{figure}[tbp]
\includegraphics[width=\columnwidth]{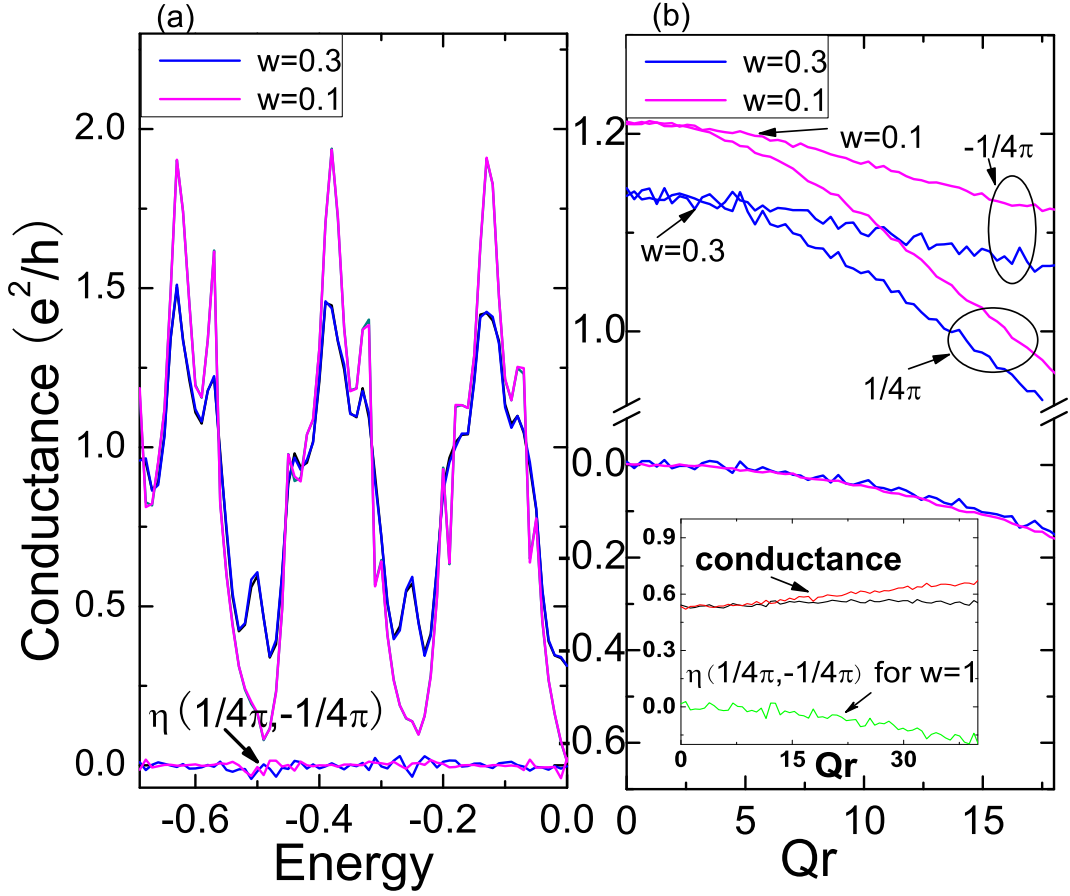}
\caption{(Color online)(a) The conductance of a 1D ring as a function of
Fermi energy $E_{F}$ without SOIs for random width $w=0.1,0.3$; (b) The
conductance and $\protect\eta $ of a 1D ring as a function of the strength
of the RSOI, for outgoing lead located at $\protect\phi =\frac{1}{4}\protect%
\pi ,-\frac{1}{4}\protect\pi $, and $Q_{r}=Q_{d}$, $E_{F}=0.1$, $w=0.1,0.3$.
The inset shows the conductance and the anisotropic ratio $\protect\eta $
when $w=1$.}
\label{fig:random}
\end{figure}

In Fig.~\ref{fig:random}(a), we plot the conductance as a function of Fermi
energy $E_{F}$ without RSOI. The ratio $\eta(\frac{1}{4}\pi, -\frac{1}{4}\pi)
$ is negligible for (weak and strong) different disorders $w=0.1,0.3$ when
the system is without the RSOI. Fig.~\ref{fig:random}(b) plots the
conductance of a 1D ring as a function of the strength of RSOI and DSOI,
when $Q_{r}=Q_{d}$, for the various random widths $w=0.1, 0.3, 1$ ($w=1$ for
inset). It can be clearly seen that the disorder-averaged conductance for
the strong disorder case ($w=1$) shows almost the same anisotropy as that
for the weak disorder case ($w=0.1,0.3$). (see Fig.~\ref{fig:random}(b))

While the anisotropy of the 1D ring becomes significant as the strengths of
the RSOI and DSOI increase, random disorder increases the scattering of the
ring, and decreases conductance compared to that of a clean 1D ring. The
anisotropic spin transport can still survive even in the presence of weak
and strong disorder.

\subsection{\label{P}The spin polarization of current}

The spin polarization vector of current $\boldsymbol{P}=(P_{x},P_{y},P_{z})$
can be evaluated as follows~\cite{Souma,BK}:
\begin{equation}
\boldsymbol{P}^{\sigma}=Tr_{s}[\hat{\rho}^{\sigma}\mathbf{\hat{\sigma}}],
\end{equation}
where the density matrix is given by:
\begin{equation}  \label{polari}
\hat{\rho}^{\sigma}=\frac{e^{2}/h}{G^{\uparrow \sigma }+G^{\downarrow \sigma}%
}\sum_{p,p^{^{\prime}}=1}^{M} \left(
\begin{array}{cc}
\vert \mathbf{t}_{p p^{^{\prime}},\uparrow \sigma} \vert^{2} & \mathbf{t}%
_{pp^{^{\prime}},\uparrow \sigma}\mathbf{t}^{*}_{pp^{^{\prime}},\downarrow
\sigma} \\
\mathbf{t}_{p p^{^{\prime}},\downarrow \sigma} \mathbf{t}^{*}_{p
p^{^{\prime}},\uparrow \sigma} & \vert \mathbf{t}_{pp^{^{\prime}},\downarrow
\sigma} \vert^{2}%
\end{array}
\right) ,
\end{equation}
where $Tr_{s}$ denotes the trace in the spin Hilbert space. Then, the spin
polarized vector $\boldsymbol{P}$ is~\cite{Souma}:
\begin{eqnarray}
P^{\sigma}_{x} & = & \frac{G^{\uparrow\sigma}-G^{\downarrow\sigma}}{%
G^{\uparrow\sigma}+G^{\downarrow\sigma}} , \\
P^{\sigma}_{y} & = &\frac{2e^{2}/h}{G^{\uparrow \sigma }+G^{\downarrow
\sigma}}\sum_{p,p^{^{\prime}}=1}^{M}\mathbf{Re}[\mathbf{t}%
_{pp^{^{\prime}},\uparrow \sigma}\mathbf{t}^{*}_{pp^{^{\prime}},\downarrow
\sigma}] , \\
P^{\sigma}_{z} & = &\frac{2e^{2}/h}{G^{\uparrow \sigma }+G^{\downarrow\sigma}%
}\sum_{p,p^{^{\prime}}=1}^{M}\mathbf{Im}[\mathbf{t}_{pp^{^{\prime}},\uparrow
\sigma}\mathbf{t}^{*}_{pp^{^{\prime}},\downarrow \sigma}] ,
\end{eqnarray}
where the $x$-axis is chosen as the spin-quantized axis, $\hat{\sigma}%
_{x}\vert\uparrow\rangle=+\vert\uparrow\rangle$ and $\hat{\sigma}%
_{x}\vert\downarrow\rangle=-\vert\downarrow\rangle$, so that Pauli spin
matrix has the following form:
\begin{equation}
\hat{\sigma}_{x}=\left(
\begin{array}{cc}
1 & 0 \\
0 & -1%
\end{array}%
\right), \ \hat{\sigma}_{y}=\left(
\begin{array}{cc}
0 & 1 \\
1 & 0%
\end{array}%
\right), \ \hat{\sigma}_{z}=\left(
\begin{array}{cc}
0 & i \\
-i & 0%
\end{array}%
\right).
\end{equation}

For the spin polarized injection, i.e., $P_{x}=1$, the magnitude of the spin
polarization $P$ in the outgoing lead will not change, i.e., $|P|=1$ since
there is no other orbit channel to interact with the spin.\cite{Zhai}

\begin{figure}[tbp]
\includegraphics[width=\columnwidth]{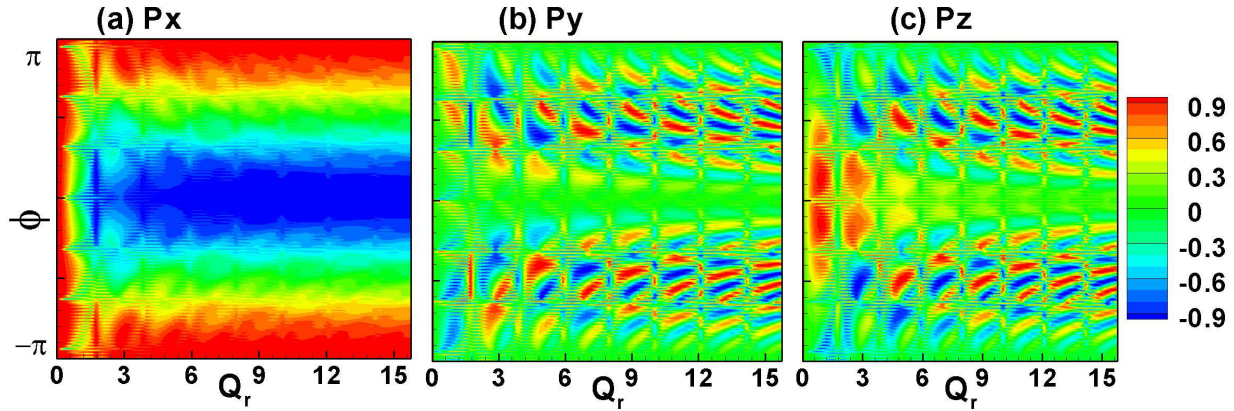}
\caption{ (Color online) The contour plot of the spin polarization of
current as a function of the strength of the RSOI $Q_{r}$ alone and the
position of the right lead in the absence of the DSOI, $E_{F}=-0.1$, $%
Q_{d}=0 $. (a) for $P_{x}$; (b) for $P_{y}$; (c) for $P_{z}$. The
spin-quantized axis is the $x$-axis.}
\label{fig:pa}
\end{figure}

Fig.~\ref{fig:pa} depicts the current spin polarization $P_{i}(i=x,y,z)$ of
a 1D ring as a function of the strength of the RSOI $Q_{r}$ and the
positions of the outgoing lead. The RSOI behaves like an effective in-plane
momentum-dependent magnetic field, and the fully spin-up polarized current
in the incoming lead will be changed to the spin-down current in the
outgoing lead at large RSOI. The three components of the outgoing
polarization vector also show cylindrical symmetry for the RSOI or DSOI
alone, since the RSOI or DSOI alone does not break the cylindrical symmetry
of a 1D ring. The spin polarization $P_{x}$ decreases rapidly from $P_{x}=1$
to $P_{x}\approx -1$ as the strength of the RSOI increases when the outgoing
lead is located at the position near $\phi =0$, while the spin polarization $%
P_{y}$ and $P_{z}$ oscillate and decrease to zero. When the outgoing lead
locates away from the $x$-axis, i.e.,$\phi =0$, $P_{y}$ and $P_{z}$
oscillate quickly with increasing $Q_{r}$.

\begin{figure}[tbp]
\includegraphics[width=\columnwidth]{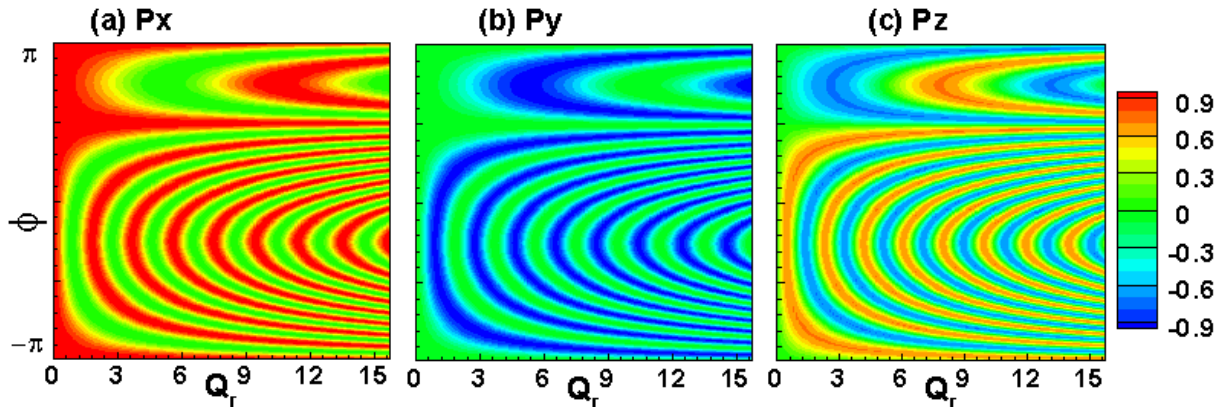}
\caption{ (Color online) The same as Fig.~\protect\ref{fig:pa}, but includes
the DSOI.}
\label{fig:pab}
\end{figure}

In Fig.~\ref{fig:pab}, we show how the spin polarizations $P_{i}(i=x,y,z)$
vary with the strength of the SOIs and the position of the outgoing lead $%
\phi$ in the presence of equal-strength RSOI and DSOI, i.e., $Q_{r}=Q_{d}$.
All three components $P_{x}$, $P_{y}$, and $P_{z}$ oscillate regularly as
the strengths of the RSOI and DSOI increase, and show significant anisotropy
of spin polarization with respect to the position of the outgoing lead. This
feature can also be understood from the interplay between the effective
periodic potential induced by the SOIs and the quantum interference. For a
fixed strength of the SOI, the asymmetric characteristic of the polarization
$\mathbf{P}$ as a function of the angle $\phi $ arises from the cylinder
symmetry breaking induced by the effective potential $\frac{\alpha\beta}{2}%
\sin{2\phi}$. The quantum interference between the spin -up and -down
electrons traveling clockwise and/or counterclockwise along the ring's upper
and lower arms leads to the oscillation of the polarization $\mathbf{P}$ as
a function of the strengths of the SOIs at a fixed angle $\phi$. Compared to
Fig.~\ref{fig:pa}, the spin polarization $P_{x}$ will decrease to $0$
instead of $-1$ as the strengths of the SOIs increase. This is because the
DSOI behaves like a twisted in-plane magnetic field, while the effective
magnetic field induced by the RSOI always points along the radial of the
ring.

\section{\label{sec:conclusions}Conclusion}

We investigate theoretically the spin transport through a two-terminal
mesoscopic ring in the presence of both the RSOI and DSOI. We find that the
interplay between the RSOI and DSOI leads to the anisotropic transport
through a two-terminal cylindrical mesoscopic ring, i.e., breaks the
cylindrical symmetry. This interesting feature arises from the periodic
potential along the ring caused by the interplay between the RSOI and DSOI.
This interplay also results in a significant variation in electron density
and the spin polarization of current. The anisotropy of the spin transport
through the mesoscopic ring induced by the interplay between the RSOI and
DSOI can survive even in the presence of the disorder effect. Furthermore,
the anisotropy of the spin transport should play an important role in the
potential application of all-electrical spintronic devices.

\begin{acknowledgments}
This work is partly supported by NSFC Grant No. 62525405 and the knowledge
innovation project of CAS.
\end{acknowledgments}

\end{document}